\documentclass[prl,twocolumn,groupedaddress,superscriptaddress,showpacs,floatfix]{revtex4-1}
\usepackage{amsfonts}
\usepackage{graphicx}
\usepackage{dcolumn}
\usepackage{bm}
\usepackage{amsmath}
\usepackage{amssymb}
\usepackage{revsymb4-1}
\usepackage{hhline}

\usepackage{siunitx}
\usepackage[normalem]{ulem}
\usepackage{color}
\definecolor{darkerblue}{rgb}{0,0,0.75}
\definecolor{darkerred}{rgb}{0.8,0,0}
\usepackage[colorlinks=false,citecolor=darkerblue,linkcolor=darkerred]{hyperref}

\newcommand{\addMisha}[1]{\textcolor{blue}{#1}}

\begin{document}
% \large

\title{Magnetic control of polariton spin transport}

\author{D. Caputo}
\affiliation{CNR-NANOTEC, Istituto di Nanotecnologia, 
  Via Monteroni, 73100 Lecce, Italy}
\affiliation{University of Salento, Via Arnesano, 73100 Lecce, Italy}

\author{E. S. Sedov}
\affiliation{School of Physics and Astronomy, 
  University of Southampton, SO17 1NJ Southampton, United Kingdom}
\affiliation{Department of Physics and Applied Mathematics, 
  Vladimir State University named after A. G. and N. G. Stoletovs, 
  Gorky str. 87, 600000, Vladimir, Russia}

\author{D. Ballarini}
\affiliation{CNR-NANOTEC, Istituto di Nanotecnologia, 
  Via Monteroni, 73100 Lecce, Italy}

\author{M.M. Glazov}
\affiliation{Ioffe Institute, 26 Polytechnicheskaya,
  194021 St. Petersburg, Russia}
\affiliation{Spin Optics Laboratory, St. Petersburg State University, Ul'anovskaya 1, Peterhof, St. Petersburg 198504, Russia}

\author{Alexey~Kavokin}
\affiliation{Institute of Natural Sciences, Westlake Institute for Advanced Study, Westlake University, Hangzhou, China}

\author{D. Sanvitto}
\affiliation{CNR-NANOTEC, Istituto di Nanotecnologia, 
  Via Monteroni, 73100 Lecce, Italy}
\affiliation{INFN, Sez. Lecce, 73100 Lecce, Italy}

\begin{abstract}
  We show the full control of the polarization dynamics of a propagating 
  exciton-polariton condensate in a planar microcavity by using a magnetic field 
  applied in the Voigt geometry.  The change of the spin-beat frequency, the 
  suppression of the optical spin Hall effect and the rotation of the polarization 
  pattern by the magnetic field are theoretically reproduced by accounting for the 
  magneto-induced mixing of exciton-polariton and dark, spin forbidden, exciton states.
\end{abstract}

%\date{\today}

\maketitle

%==================================================
%==================================================
%==================================================
%==================================================
%==================================================
%==================================================
%==================================================
%==================================================

%\section{Introduction}
The remarkable progresses in the control of matter-light interaction 
in semiconductor optical microcavities have made it possible to design 
a new generation of optoelectronic 
devices~\cite{PhysLettA2141931996,PhysRevB.85.235102,
  NatureCommun417782013,PhysRevLett.101.266402,PhysRevB.81.125327,
PhysRevLett.102.046407,  NaturePhotonics43612010}.
These are based on the peculiar properties of exciton-polaritons, 
half-light half-matter bosonic quasiparticles 
arising from the strong coupling between photonic cavity modes and
excitons in quantum wells placed inside the cavity. 
One of the most remarkable properties of polaritons is 
that they have a spin degree of freedom inherited from the photon 
chirality and exciton spin angular momentum that shows long coherence 
time and the possibility to be actively manipulated by external fields 
through the excitonic component~\cite{PhysStatusSolidiB24222712005, PhysRevB.95.085429}.
This additional feature of polaritons significantly broadens 
the range of their possible applications to include what is known 
under the name of  \textit{spinoptronics}.
By now, there
are already several implemented concepts in the form of
spinoptronics devices such as the
``Datta and Das'' spin transistor~\cite{PhysRevB.81.125327, PhysRevB.87.195305},
the polaritonic analogue of a Berry-phase interferometer~\cite{PhysRevLett.102.046407}
and the exciton-polariton spin switch \cite{NaturePhotonics43612010, Ballarini2009}.
In all these works a central aspect is the ability to control the spin of polaritons 
using internal as well as external factors to affect their 
polarization properties.

From a fundamental point of view, the dominant effect on the polariton 
spin dynamics is the optical spin Hall effect 
(OSHE)~\cite{PhysRevLett.95.136601,PhysRevLett.109.036404,PhysRevBAssmann}.
Basically, it originates from the longitudinal-transverse (LT) splitting 
of exciton polariton states. 
The effect of this splitting can be described by an effective magnetic field,
strongly dependent on the direction of the quasi-particle propagation
and its velocity, giving rise to the polariton
spin precession as it propagates in the cavity plane.
One of the main problem, in particular when polarization is a key parameter in 
polariton devices, is the control of such an effective magnetic field which in 
turn directly affects the polariton state.
Usually, the Faraday geometry with the magnetic field directed along the cavity
normal is used. In this configuration the studies of the influence of the 
magnetic field on the 
polariton dispersion~\cite{PhysRevB.91.075309}, 
coherence properties~\cite{Chernenko2016} 
as well as on the spin textures in excitonic~\cite{PhysRevLett.110.246403} and 
polaritonic~\cite{PhysRevB.88.035311} systems have been reported.
Hovewer, such geometry cannot be effective on the control of the OSHE~\cite{Rubo2006227,PhysRevLett.105.256401,PhysRevLett.106.257401,SciRep797972017}.

By contrast, the effect on exciton-polariton spin dynamics 
in the Voigt geometry, 
with the external magnetic field within the plane of the quantum well, has
not been studied so far. This is because the in-plane
field does not directly couple with the polariton pseudospin 
and one may expect its effect to be quite minor.
In this Letter we demonstrate that this is not the
case by reporting for the first time the direct control of the 
polariton spin transport both in a confined one-dimensional (1D) geometry
and in the whole two-dimensional (2D) plane of the cavity through the 
application of an external magnetic field directed in the cavity plane.
In this context, we show the possibility to control and even totally 
suppress the OSHE for polaritons propagating in a given direction 
by properly choosing the magnitude and the direction of the applied field.
Moreover the application of the external
magnetic field causes a stretching of the circular pseudospin patterns 
in the axis normal to the magnetic field direction and a contraction in the 
same direction of the external field.
From the point of view of applications, the possibility
 to completely control the OSHE related intensity 
oscillations allows one to remove the residual
density modulations that appear during polaritons propagation
 and that would be detrimental for the elaboration of the signal 
inside the devices \cite{PhysRevB.87.195305}.
All the experimentally observed phenomena can be described within an analytical 
model taking into account both the field-induced mixing of bright polariton 
doublet with dark excitonic states
and nonlinear effects originating from the polariton-polariton interactions.

%==============================================
%ffffffffffffffffffffffffffffffffffffffffffffff
%iiiiiiiiiiiiiiiiiiiiiiiiiiiiiiiiiiiiiiiiiiiiii
%gggggggggggggggggggggggggggggggggggggggggggggg
\begin{figure}%[b]
  \centering
  \includegraphics[width=1\columnwidth]{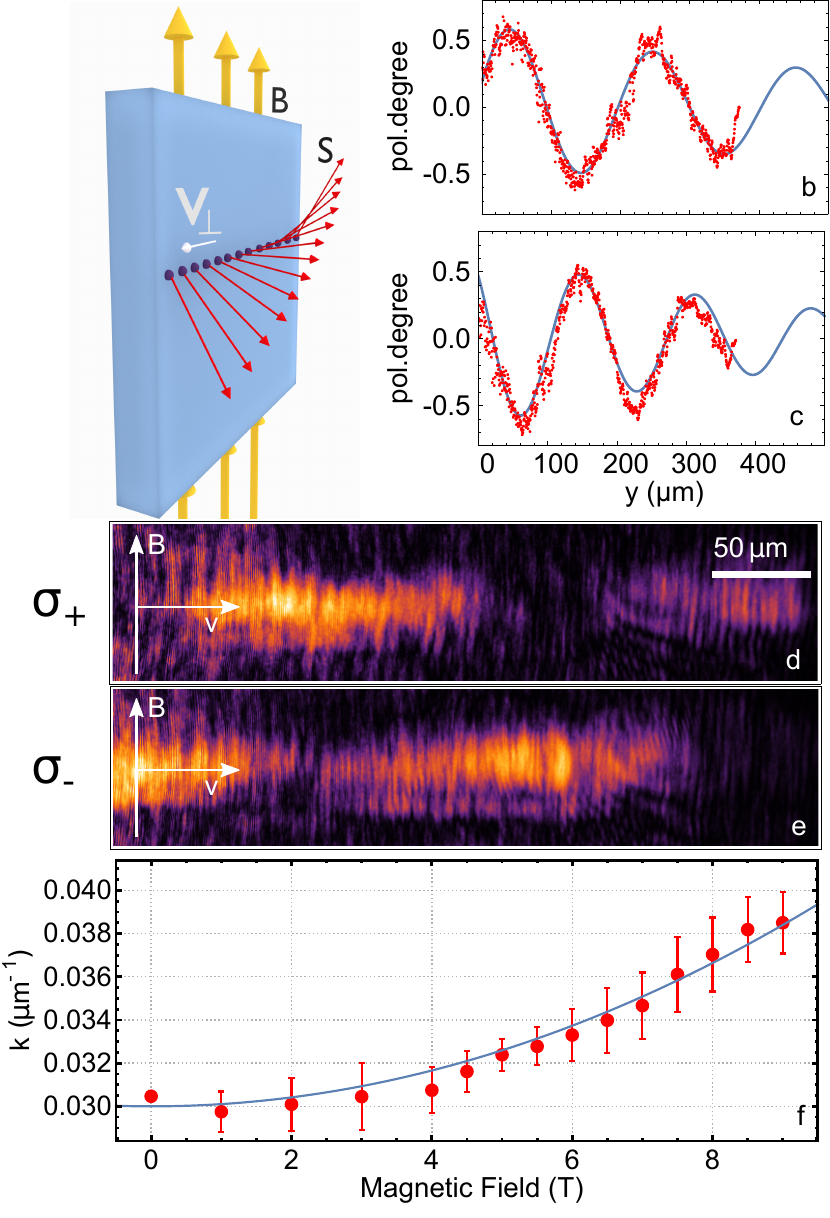}
  \caption{
    (a) Sketch of the experimental setup with the external
    magnetic field ($\mathbf B$) in the microcavity plane (Voigt configuration). The polariton propagation direction is perpendicular to the field, $\mathbf v \perp \mathbf B$.
    (b,c) Circular polarization degree vs. $y$-coordinate for $B=0$~T (b) and $B=9$~T (c), $y=0$ corresponds to the distance of 50~$\mu$m from the excitation spot, smaller distances are removed to avoid contribution of the scattered light.
    Red curves show experimental data, blue lines are fits after Eq.~\eqref{SzSolutionGeneral}.
    (d,e) False color plots of spatial distribution of intensities in $\sigma^+$ (co-polarized, d) and $\sigma^-$ (e).
    (f) propagation constant $\kappa$ in Eq.~\eqref{SzSolutionGeneral} as a function of $B$ (dots are experiment, solid curve is the fit after Eq.~\eqref{PropConst} with
    $v=1.52~\mu\text{m/ps}$,
    $\hbar \Delta_{\text{LT}} = 30\,\mu$eV,
    $\hbar \beta = 0.1 \,\mu\text{eV/T}^2$.
      }
  \label{fig:1}
\end{figure}
%==============================================
%ffffffffffffffffffffffffffffffffffffffffffffff
%iiiiiiiiiiiiiiiiiiiiiiiiiiiiiiiiiiiiiiiiiiiiii
%gggggggggggggggggggggggggggggggggggggggggggggg

%==============================================
%ffffffffffffffffffffffffffffffffffffffffffffff
%iiiiiiiiiiiiiiiiiiiiiiiiiiiiiiiiiiiiiiiiiiiiii
%gggggggggggggggggggggggggggggggggggggggggggggg
\begin{figure}%[htbp]
  \centering
  \includegraphics[width=0.9\columnwidth]{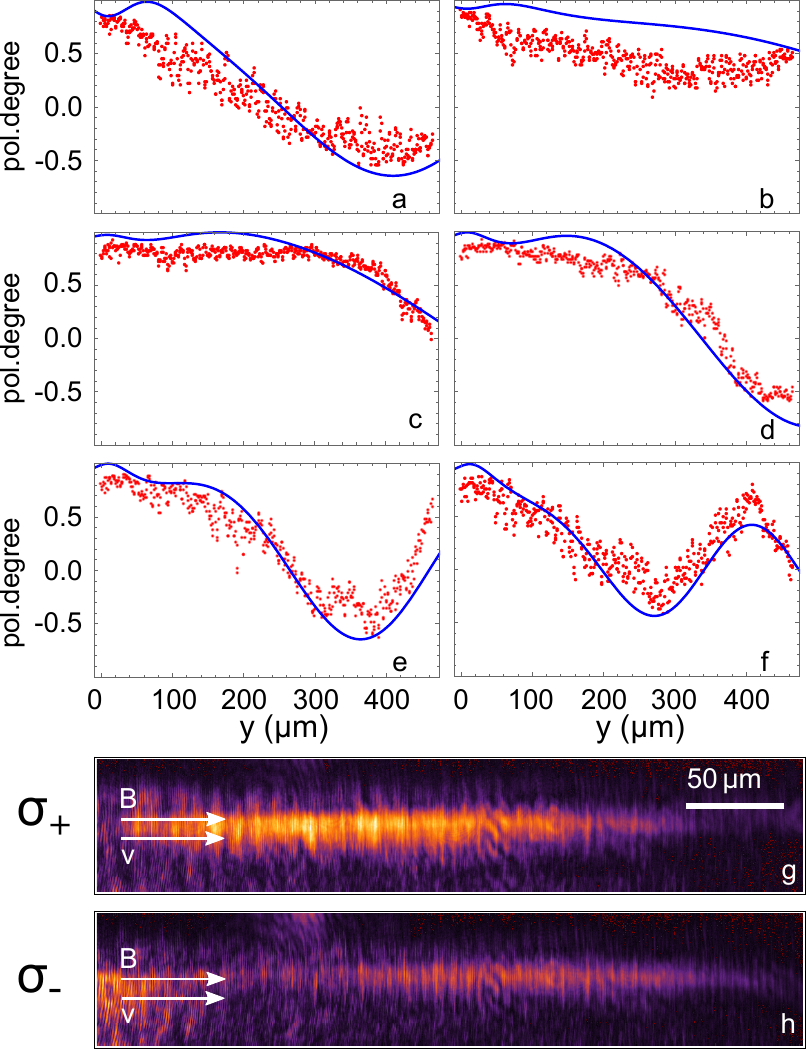}
  \caption{
  (a)--(f) $P_c(y)$ for the magnetic field oriented along the propagation direction $\mathbf B \parallel y$. Magnetic field values are B=0 T, 3 T, 5.5 T, 6.5 T, 7.5 T, 8.5 T from (a) to (f), respectively. 
(g,h) False color plots of spatial distribution of intensities in $\sigma^+$ (co-polarized, g) and $\sigma^-$ (cross-polarized, h).
In the panels (a)--(f) red dots are experiment, blue solid curves are solutions of the nonlinear equation~\eqref{PseudospinEqMotion} with the following parameters:
    $v = 0.35\, \mu \text{m/ps}$,
	$\hbar  \Delta_{\text{LT}} = 2.5\,\mu$eV,
	$\hbar \alpha S_0 = 30\, \mu$eV,
	$\gamma _{s} = 0.004\,\text{ps}^{-1}$,
	$\hbar \beta = 0.1 \,\mu\text{eV/T}^2$.
	The initial conditions are $S_{z_0}=0.9$, $S_{y_0}=-0.15$, $S_{x_0}=-(1-S_{y_0}^2-S_{z_0}^2)^{1/2}$.
  }
  \label{fig:1b}
\end{figure}
%==============================================
%ffffffffffffffffffffffffffffffffffffffffffffff
%iiiiiiiiiiiiiiiiiiiiiiiiiiiiiiiiiiiiiiiiiiiiii
%gggggggggggggggggggggggggggggggggggggggggggggg

%\section{Results}

The sample studied in this work is a high finesse 3$\lambda$/2  
GaAs/AlGaAs planar microcavity grown along $z\parallel [001]$ axis 
with a state-of-the-art polariton lifetime of about 100~ps. The high quality factor $Q>{10^5}$ and the low density of defects allow ballistic propagation of polaritons to cover several hundreds of micron, as recently demonstrated by different groups~\cite{Caputo2017, PhysRevB.88.235314}.
The microcavity contains 12 GaAs quantum wells (7~nm-wide) placed at three 
anti-node positions of the electric field, providing a vacuum Rabi splitting of
$\SI{16}{\milli \electronvolt}$.
The front (back) mirror consists of 34 (40) pair of AlAs/Al$_{0.2}$Ga$_{0.8}$As layers 
and cavity-exciton detuning is slightly negative, about 
$\SI{-2}{\milli \electronvolt}$. 
Polaritons are injected both resonantly and nonresonantly using a low-noise, 
narrow-linewidth Ti:sapphire laser with stabilized output frequency in a 
continuous wave operation mode. The sample is kept at a temperature of around 10 K.
The sample emission is collected, filtered in polarization and sent to the 
entrance slit of a spectrometer coupled with a charge coupled device camera.

%==============================================
%==============================================
%==============================================
%RRRRRRRRRRRRRRRRRRRRRRRRRRRRRRRRRRRRRRRRRRRRRR
%EEEEEEEEEEEEEEEEEEEEEEEEEEEEEEEEEEEEEEEEEEEEEE
%SSSSSSSSSSSSSSSSSSSSSSSSSSSSSSSSSSSSSSSSSSSSSS
%==============================================
%==============================================
%==============================================

In order to study the 1D case, 
we resonantly inject $\sigma^+$ circularly polarised polaritons with a 
finite momentum into natural misfits dislocations present along the $[1\bar10]$
axis of the sample.
The 1D confinement is shallow but allows to observe the polariton propagation 
up to 400~$\mu \text{m}$ from the laser injection point with a negligible
spread in the perpendicular direction (see Fig.~\ref{fig:1}).
The coherent spin oscillations during the propagation of polaritons are 
measured by selectively detecting the emission intensity co- ($\sigma^+$) and 
cross-polarised ($\sigma^-$) 
with respect to the exciting laser, as shown in Fig.~\ref{fig:1}d and Fig.~\ref{fig:1}e, respectively.
The circular polarization degree, $P_{\rm c}$, is then obtained as 
$P_{\rm c}={(I_{\sigma^{+}}-I_{\sigma^{-}})}/{(I_{\sigma^{+}}+I_{\sigma^{-}})}$. 
For data shown in Fig.~\ref{fig:1}, polaritons are injected with a speed of 
$1.5 \, \mu \text{m/ps}$ and the magnetic field, applied in the plane of 
polariton propagation normally to the propagation direction, 
spans a range from $0$ to $9$~T. 
The pronounced oscillations of the circular polarization degree
as a function of coordinate are observed being a signature of 
the polariton pseudospin precession in the course
of propagation. Interestingly and somewhat unexpectedly, 
the magnetic field affects  the frequency of 
the spatial oscillations in $P_{\rm c}$ that increases 
quadratically with the intensity of the magnetic field, as shown 
in Fig. \ref{fig:1}(g).
Figure \ref{fig:1b} shows the results of measurement performed with the 
sample rotated by $90^{\circ}$ with respect to the orientation shown 
in Fig.~\ref{fig:1}, 
in order to have the external magnetic field oriented parallel to the 
dislocation and, hence, to the polariton propagation direction.
A deeper dislocation is considered in this case, showing a non-trivial 
polarization pattern along propagation.
Indeed, increasing the magnetic field intensity, the spatial oscillations frequency 
first decreases (up to 5~T) and then increases with a constant positive offset 
in the co-polarised component.

The observed effects can be quantitatively described within a pseudospin 
model parametrizing the polariton spin density matrix through the vector 
$\mathcal S_{\mathbf k}$ whose $z$ component describes the circular polarization 
degree of the particles and the in-plane components characterize the linear 
polarization degree in two sets of axes.
The equation of motion for the pseudospin of polaritons in the $\mathbf{k}$ 
state is written as~\cite{PhysRevLett.95.136601}
\begin{equation}
  \label{PseudospinEqMotion}
  \frac{\partial \mathbf{S}_{\mathbf{k}}}{\partial t} +  \mathbf{S}_{\mathbf{k}} \times \mathbf{\Omega}_{\mathbf k} + \gamma _{s} \mathbf{S}_{\mathbf k} =0 ,
\end{equation}
Hereafter we assume ballistic propagation of polaritons, use the set of axes with 
$x\parallel [110]$, $y\parallel [1\bar 10]$ and $z\parallel [001]$,  
$\mathbf \Omega_{\mathbf k}$ is the effective pseudospin precession frequency and 
the last term accounts for the spin relaxation processes with the rate $\gamma_s$. 
The effective precession frequency components read
\begin{subequations}
\label{Omega}
\begin{align}
\Omega_{\mathbf{k}, x} &= \left.  \Delta_{\text{LT}} (k_x^2 - k_y^2)  \right/ k^2 - \beta  (B_x^2-B_y^2),\\
\Omega_{\mathbf{k}, y} &= \left. 2 \Delta_{\text{LT}} k_x k_y   \right/ k^2  - 2\beta B_x B_y,\\
\Omega_{\mathbf{k},z} &= \alpha S_z,
\end{align}
\end{subequations}
 and 
contain contributions from the LT splitting of polariton modes, $\Delta_{\text{LT}}$~\cite{PhysRevB.59.5082}, in the linear polarization components due to the applied external magnetic field in the cavity plane $\beta B_{x,y}^2$, and the so-called self-induced Larmor precession of the polariton pseudospin due to the polariton-polariton interactions ($\alpha$) treated here within the mean-field approach~\cite{PhysStatusSolidiB24222712005,PhysRevB.77.075320,PhysRevB.91.161307}.
Importantly, the in-plane components of the effective field contain the 
\emph{quadratic} magnetic field contributions. The form of these terms follows from the symmetry arguments since the quadratic combinations $B_i B_j$ and $k_i k_j$ with $i,j=x,y$ transform in the same way already in the isotropic approximation and the effects of $C_{2v}$ point symmetry of the studied structure on the magnetic-field induced terms are disregarded. Microscopically, the parameter $\beta$ results from magneto-induced mixing of polariton states and dark (spin-forbidden) excitons with the additional contribution from the diamagnetic effect, just like for excitons in quantum wells and quantum dots~\cite{PhysRevB.73.033306,PhysRevB.76.193313}, see Supplementary material~\cite{SupplRef} for the 
details.

Equation~\eqref{PseudospinEqMotion} can be solved analytically for various experimentally relevant configurations. Let us first set $\alpha=0$, i.e., neglect the effect
of the polariton-polariton interactions and assume that the polaritons propagate along the $y$-axis and $\mathbf B$ is either parallel or perpendicular to the polariton velocity.
From Eqs.~\eqref{Omega} it follows that $\Omega_y\equiv 0$ and
\begin{equation}
\label{SzSolutionGeneral}
S_z (y) = S_{z_0} \cos \left( \kappa y  \right) e^{-\left. y \right/ \ell _s}\addMisha{.}
\end{equation}
Here the propagation constant of the oscillatory distribution of pseudospin is:
\begin{equation}
\label{PropConst}
\kappa = \left.  \left[ {\Delta_{\text{LT}} + \beta  (B_x^2-B_y^2)}  \right] \right/v,
\end{equation}
and the pseudospin decay length $\ell _s= v/\gamma_{s}$, where
$\mathbf v=  \mathbf k/m$ (with $m$ being the polariton effective mass) is 
the polariton velocity.
For  positive $\beta$  and $\mathbf{B} \perp \mathbf{k}$, Eq.~\eqref{PropConst} predicts a 
monotonic increase of the propagation constant $\kappa$ with the square of the magnetic
 field magnitude $B^2$, in full agreement with the experimental results shown in 
Fig.~\ref{fig:1}(g).
By contrast, with $\mathbf{B} \parallel \mathbf{k}$ the dependence $\kappa(B)$ is more complicated: from Eq.~\eqref{PropConst}, it follows that $\kappa$ is a nonmonotonic function of $B$. First it decreases for $B< B_c = \sqrt{ \Delta_{\text{LT}}  /\beta}$ and 
then increases for higher field intensities.
Under the condition $B = B_c$ the complete stop of oscillations is expected due to the suppression of the LT-splitting by the magnetic field similarly to the field-induces suppression of exciton anisotropic splitting in quantum dots~\cite{PhysRevB.73.033306,PhysRevB.76.193313}. Qualitatively, these results are in agreement with the experimental data presented in Fig.~\ref{fig:1b}, with $B_{c}\approx\SI{4}{\tesla}$.

% ==============================================
% ffffffffffffffffffffffffffffffffffffffffffffff
% iiiiiiiiiiiiiiiiiiiiiiiiiiiiiiiiiiiiiiiiiiiiii
% gggggggggggggggggggggggggggggggggggggggggggggg
\begin{figure}
  \centering
  \includegraphics[width=0.9\columnwidth]{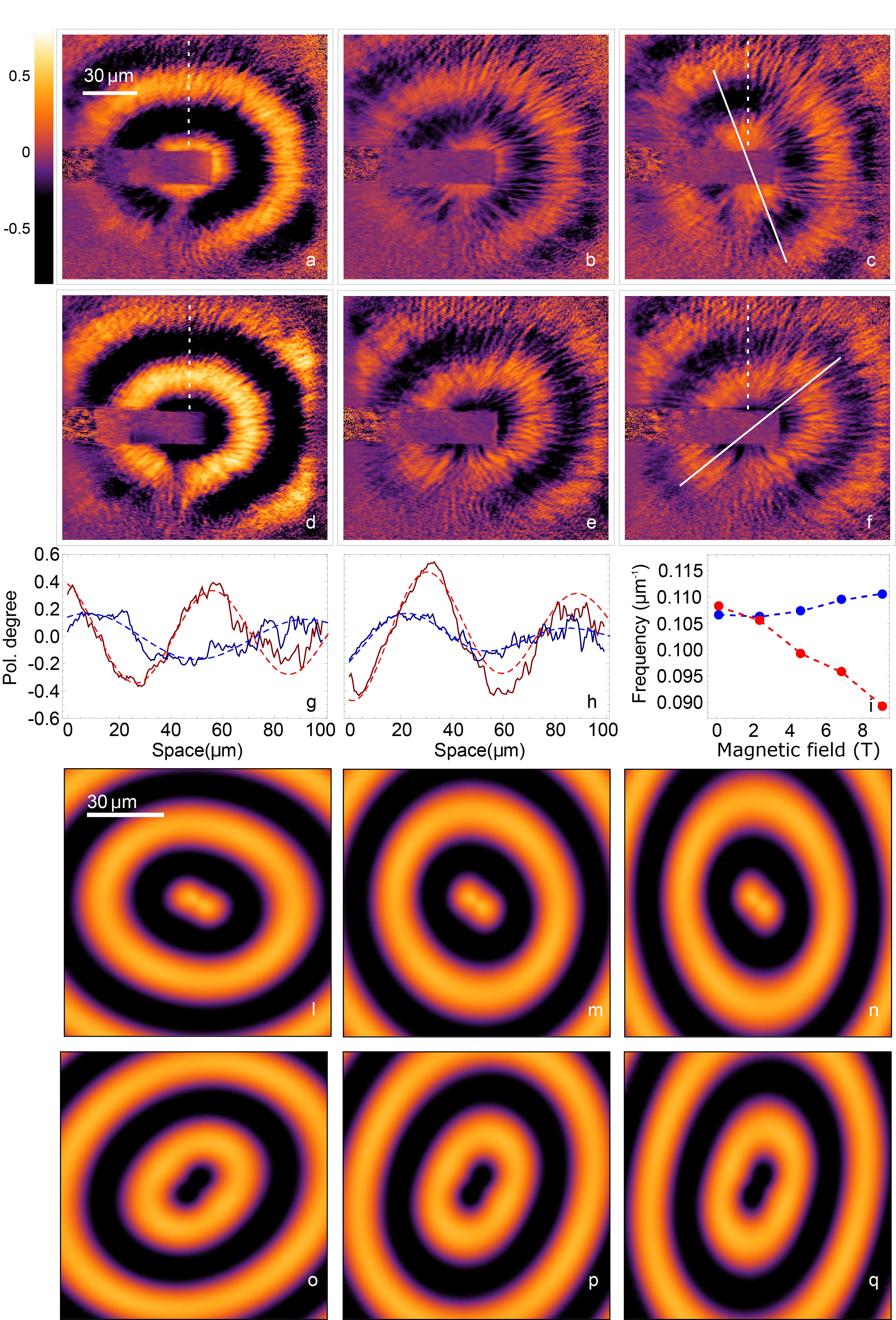}
  \caption{
  Measured circular polarization degree patterns with $\mathbf{B}\parallel y$ 
    for the magnetic field intensities of $(0, 7, 9)$~T 
    and the initial polarizations $\sigma^{+}$ (a,b,c) 
    and $\sigma^{-}$ (d,e,f).
    The propagation velocity is about 1.8~$\mu \text{m/ps}$.
    (g) Cross section of $S_{z}$ along the vertical 
    directions at 0~T (red line) and 9~T (blue line) 
    as indicated by the dashed line in (a) and (c), respectively.
    The fitting function is $Ae^{-by} \sin(\kappa y+\phi)$.
    (h) Same as in (g) but for $\sigma^{-}$ (see (d) and (f)).
    (i) propagation constant $\kappa$ extracted as best fit to 
    the data for $\sigma^{+}$ (red) and $\sigma^{-}$ 
    (blue) polarization.
    (l)--(q) The circular polarization degree in real space simulation based on Eq.~\eqref{PseudospinEqMotion}.
    The magnetic field intensities correspond to ones in (a-f).
    Values of the parameters used in the model are following:
    $\hbar \Delta _{\text{LT}} = 250\, \mu \text{eV}$,
    $\hbar \beta=0.8 \, \mu \text{eV/T}^2$,
    $\alpha S_{0} = 30\, \mu \text{eV}$,
    $\gamma _s = 0.004 \,\text{ps}^{-1}$.
    The initial conditions are $S_{x_0}=-0.4$, $S_{y_0}=(1-S_{x_0}^2-S_{z_0}^2)^{1/2}$, $S_{z_0} = 0.9$ and  $S_{z_0} = -0.9$ for (l--n) and (o--q), respectively .
   }
  \label{fig:2}
\end{figure}

While Eqs.~\eqref{SzSolutionGeneral} and \eqref{PropConst} quantitatively fit the data in Fig.~\ref{fig:1}, see solid lines in the panels (b,c), the linear model is not sufficient to describe all the peculiarities  of the 
polariton polarization dynamics shown in Fig.~\ref{fig:1b}. This is because the nonlinearity due to spin-dependent polariton-polariton interactions becomes of particular importance in the situation of $B\approx B_c$. Also, in the experimental geometry with $\mathbf B \parallel \mathbf k$, the polariton propagation velocity is 
relatively small, $v \approx 0.3 \, \mu\text{m} / \text{ps}$, that results in the weaker 
manifestation of the LT-splitting effect. 
Indeed, one of the peculiarities seen in Fig.~\ref{fig:1b} is the positive offset in the $S_{z} (y)$-dependence. 
This effect is the manifestation of the presence of the third component of the
 effective magnetic field  $\Omega _z \propto S_z $, which tilts the pseudospin precession axis towards the $z$-axis and suppresses partially the effect of the LT-splitting~\cite{PhysRevB.88.035311, 
ApplPhysLett1100611082017}.
The results of calculations for this configuration are shown by blue curves in Fig.~\ref{fig:1b} and closely match the experimental points.
Note that, as expected, the self-induced Zeeman splitting due to the 
polariton-polariton interactions decreases at lower densities and
the dynamics of $S_z(y)$ returns to the 
harmonic character at the distance of about 300~$\mu$m, Fig.~\ref{fig:1b}.

% ==============================================
% ==============================================
% ==============================================
% 2222222222222222222222222222222222222222222222222
% DDDDDDDDDDDDDDDDDDDDDDDDDDDDDDDDDDDDDDDDD
% ==============================================
% ==============================================
% ==============================================

In order to understand how the complete two dimensional polarization spatial 
distribution is affected by the application of an in-plane magnetic field,
we performed a different experiment using a free, high velocity, radially 
expanding polariton condensate in a clean region of the sample.
Even though obtained with a nonresonant circularly polarized 
pumping scheme, polaritons, in our case, inherit about $40 \%$ of circular 
polarization from the pump~\cite{PhysRevLett.109.036404}.
With the aim of efficiently injecting polaritons, the energy of 
the pump is tuned at the first minimum of the reflection stop band 
making it possible to reach the condensation density threshold in a region 
within the laser spot, blueshifted of about 4~meV from the bottom of the 
lower polariton branch. From this central region, a polariton flow is 
ballistically expelled and is free to radially propagate in the plane of the cavity
outside the excitation spot region, with an acceleration due to the gradient 
in the potential resulting from the blueshift under the excitation spot~\cite{PhysRevLett.118.215301}.
This configuration enables polaritons to ballistically propagate with a speed of 
about 2~$\mu \text{m/ps}$.

Figure~\ref{fig:2}(a--f) shows the $P_c$ for different magnetic field intensities.
A small asymmetry in the distribution function already present at $B=0$~T is initially compensated, 
and then enhanced by the external magnetic field, see 
Fig.~\ref{fig:2}(a--c). Indeed, the pattern was at $B=0$~T elliptical 
with the horizontal axis greater than the vertical but, by increasing the applied
magnetic field, these axes become inverted and the ellipse gets rotated. 
This is confirmed
by the opposite behaviour of the spatial frequency respect to the vertical and 
horizontal axes as reported in the inset of Fig.~\ref{fig:2}.
By changing the polarisation degree of the 
exciting laser from right-circular to left-circular, we change the 
relative orientation of the polariton spin with respect to the magnetic 
field. In Fig.~\ref{fig:2}(d--f), we show that in this case the long 
eccentricity axis at $9$~T is oriented roughly perpendicularly to 
Fig.~\ref{fig:2}(c).

%\section{Discussion}

The resulting circular polarization degree of polaritons during the expansion in the two-dimensional plane can be qualitatively described by the same equation~\eqref{PseudospinEqMotion} applied to all $\mathbf{k}$ 
on the elastic circle of the radius $k$.
Figure~\ref{fig:2}{(l--q)}  illustrates the 2D expansion of the $P_c$ theoretically 
predicted from the model.
The values of the effective magnetic field magnitude $B$ in this figure correspond to those in Fig.~\ref{fig:2}(a)--(f).
In order to distinguish the influence of different effects on the shape of 
the polarization pattern it is possible to consider the following.
At the pure circular initial polarisaton, in the presence of only the LT-splitting, the circular polarization degree patterns in real space  are rotationally invariant.
A slight tilt and the squeezing of the patterns in $x$ direction even in the absence of the external magnetic field are due to the fact that the initial polarization is not pure circular but it contains a small admixture of the linear components, see caption to Fig.~\ref{fig:2}.
Moreover, the external magnetic field tends to change the spatial frequency $\kappa$
with, as shown above, a different effect depending on the relative direction 
of the external field and the propagation.
According to Eq.~\eqref{PropConst}, the absolute value of the propagation constant $\kappa(k_y,B)$ decreases with increasing $B$ (directed along the y-axis) until 
the magnetic field compensates the effect of the LT 
splitting while a further increase in $B$ leads to the increase of 
the spatial frequency.
In the propagation direction orthogonal to B, 
the propagation constant $\kappa(k_x,B)$ monotonically increases with the 
increase of $B$ on all extent.

%CCCCCCCCCCCCCCCCCCCCCCCCCCCCCCCCCCCCCCCCCCCCCC
%OOOOOOOOOOOOOOOOOOOOOOOOOOOOOOOOOOOOOOOOOOOOOO
%NNNNNNNNNNNNNNNNNNNNNNNNNNNNNNNNNNNNNNNNNNNNNN
%CCCCCCCCCCCCCCCCCCCCCCCCCCCCCCCCCCCCCCCCCCCCCC
%LLLLLLLLLLLLLLLLLLLLLLLLLLLLLLLLLLLLLLLLLLLLLL

In conclusions, we have experimentally demonstrated the control of 
the optical spin Hall effect by tuning an external magnetic field 
applied in the direction of propagation of polaritons. 
This can be useful to avoid 
unwanted rotation of the polarization in polariton devices or by controlling
the spin degree at a given position. 
In fact, if the spin precession is instead required~\cite{PhysRevLett.109.036404}, 
the spin beat frequency can be tuned by the external magnetic field 
applied perpendicularly to the propagation direction of polaritons. 
In the 2D expansion the in-plane magnetic field induces an additional 
polarization anisotropy in the structure that manifests itself as a 
deformation of the pseudospin patterns in real space.
We have developed a theoretical model that qualitatively explains the 
observed effects.

%AAAAAAAAAAAAAAAAAAAAAAAAAAAAAAAAAAAAAAAAAAAAAA
%CCCCCCCCCCCCCCCCCCCCCCCCCCCCCCCCCCCCCCCCCCCCCC
%KKKKKKKKKKKKKKKKKKKKKKKKKKKKKKKKKKKKKKKKKKKKKK
%NNNNNNNNNNNNNNNNNNNNNNNNNNNNNNNNNNNNNNNNNNNNNN
%OOOOOOOOOOOOOOOOOOOOOOOOOOOOOOOOOOOOOOOOOOOOOO
%WWWWWWWWWWWWWWWWWWWWWWWWWWWWWWWWWWWWWWWWWWWWWW
%LLLLLLLLLLLLLLLLLLLLLLLLLLLLLLLLLLLLLLLLLLLLLL

\section*{Acknowledgments}

E.S. acknowledges support from the RFBR grant No.~16-32-60104.
A.K. and E.S. acknowledge support from the EPSRC Programme grant on Hybrid Polaritonics No. EP/M025330/1.
A.K. acknowledges the partial support from the HORIZON 2020 RISE project CoExAn (Grant No.~644076).
M.M.G. was partially supported by RF President grant MD-1555.2017.2, RFBR project 17-02-00383, SPSU and DFG Project 40.65.62.2017, and SPSU research grant 11.34.2.2012.
D.C., D.B and D.S. acknowledge the ERC project POLAFLOW (Grant N. 308136) and the ERC project ElecOpteR (grant number 780757).

\end{document}